Manuscript:

# Optimal Griffiths Phase in Heterogeneous Human Brain Networks: Brain Criticality Embracing Stability and Flexibility across Individuals

**Kejian Wu[1], Dante R. Chialvo[2,3,4], Changsong Zhou[4,5\*], and Lianchun Yu[1]\***

[1] Lanzhou Center for Theoretical Physics, Key Laboratory of Theoretical Physics of Gansu Province, Key Laboratory of Quantum Theory and Applications of MoE, Gansu Provincial Research Center for Basic Disciplines of Quantum Physics, Lanzhou University, Lanzhou 730000, China

[2] Instituto de Ciencias Físicas (ICIFI-CONICET), Center for Complex Systems and Brain Sciences (CEMSC3), Escuela de Ciencia y Tecnología, Universidad Nacional de Gral. San Martín, Campus Miguelete, 25 de Mayo y Francia, 1650, San Martín, Buenos Aires, Argentina

[3] Consejo Nacional de Investigaciones Científcas y Técnicas (CONICET), Godoy Cruz 2290, 1425, Buenos Aires, Argentina

[4] Department of Physics, Centre for Nonlinear Studies, Beijing-Hong Kong-Singapore Joint Centre for Nonlinear and Complex Systems (Hong Kong), Institute of Computational and Theoretical Studies, Hong Kong Baptist University, Kowloon Tong, Hong Kong

[5] Life Science Imaging Centre, Hong Kong Baptist University, Kowloon Tong, Hong Kong

\* Corresponding authors: Changsong Zhou (cszhou@hkbu.edu.hk)

Lianchun Yu (yulch@lzu.edu.cn).

**Abstract:**

A prominent hypothesis in neuroscience proposes that brains achieve optimal performance by operating near a critical point. However, this framework, which often assumes a universal critical point, fails to account for the extensive individual variability observed in neural dynamics and cognitive functions. These variabilities are not noise but rather an inherent manifestation of a fundamental systems-biology principle: the necessary trade-off between robustness and flexibility in human populations. Here, we propose that the Griffiths phase (GP), an extended critical regime synergically induced by two kinds of heterogeneities in brain network region and connectivity, offers a unified framework for brain criticality that better reconciles robustness and flexibility and accounts for individual variability. Using Human Connectome Project data and whole-brain modeling, we demonstrated that the synergic interplay between structural network modularity and regional heterogeneity in local excitability yields biologically viable GP featured with widely extended global excitability ranges, with an embedded optimal point that balances global/local information transmission. Crucially, an individual's position within the GP gives rise to unique global network dynamics, which in turn confer a distinctive cognitive profile via flexible configuration of functional connectivity for segregation, integration, and balance between them. These results establish GP as an evolved adaptive mechanism resolving the robustness-flexibility trade-off, fulfilling diverse cognitive demands through individualized criticality landscapes, providing a new framework of brain criticality.

**Introduction**

The "critical brain" hypothesis posits that neural systems operate near a critical point—a concept drawn from statistical physics marking the transition point between ordered and disordered phase[1, 2]. Mounting empirical evidence has bolstered this hypothesis, unveiling the pivotal role of criticality in optimizing many aspects of neural function across various levels of neural systems[3, 4, 5, 6, 7, 8, 9, 10, 11, 12]. Moreover, deviations from critical dynamics have been observed in brain disorders[13, 14]. Yet this paradigm faces a fundamental challenge: its reliance on a single, universal critical point fails to accommodate the pervasive individuality of the brain's structure, dynamics, connections, and cognitive performance[15, 16, 17, 18]. Interindividual differences in cognitive and behavioral functions are fundamentally mediated by neurobiological differences shaped by genetic, neural, and environmental factors[19, 20, 21, 22]. Such differences are potentially rooted in an evolutionary trade-off balancing robustness and flexibility in the biological population—a core principle of systems biology[23, 24]. This contradiction between the assumption of single point criticality and strong neural variability may manifest in persistent theory–experiment discrepancies regarding brain criticality[25, 26], as empirical observations show that critical exponents exhibit context-dependent heterogeneity across species, individuals, temporal domains, and stimuli[27, 28], diverging from a single set of exponents theoretically predicted by critical branching process models[29, 30] assuming a single critical point. Thus, while criticality's functional advantages remain undisputed, the monolithic critical point construct appears biologically untenable—unable to explain significant variability among healthy individual or resolve theory-experiment discrepancies.

Previous modeling studies indicate that hierarchical modular organization in brain structural connectivity (SC) networks substantially reshapes the critical brain dynamics by inducing a Griffiths phase (GP), a regime characterized by critical dynamics extending across broad parameter ranges rather than at a singular point[31, 32, 33]. While GP provides the potential to explain the context-dependent critical exponents through its parameter-dependent non-universal critical dynamics[34] and the robustness of brain function against parameter variation via its extended optimization capacity[31], it has not been explicitly shown that GP can accommodate the human brain at the individual level. This is not only a question of capturing variability in neural dynamics around criticality[35], but, more importantly, a question of accounting for personalized cognitive profiles by linking them to individual variability in brain critical dynamics. These gaps necessitate answering a fundamental question: how does individual variability originate from a brain network's GP-empowered reconciliation of robustness and flexibility? Resolving these issues is essential for developing a unified framework for brain criticality that enables brain networks to balance robustness with flexibility while employing the functional advantages of criticality to accommodate individual variability in both brain dynamics and functions.

Addressing these questions reveals two fundamental challenges. Critical dynamics have been traditionally attributed to a regulated global excitability state, sustained by a fine-tuned excitation-inhibition (E-I) balance within cortical networks[36, 37]. However, the brain is not uniform; its functional specialization is underpinned by significant regional heterogeneity in brain architecture and physiology, where variations in cyto-, myelo-, chemoarchitecture, neural receptors and gene expression give rise to distinct local E-I ratios[38, 39, 40, 41, 42]. Recent studies have demonstrated that the regional heterogeneity has great impacts on the whole brain dynamics[43, 44] and shapes empirical functional connectivity (FC) networks[43, 45]. Mechanistically, while nodal (i.e., regional) heterogeneity has been theoretically established—alongside topological modularity—as a driver of GP in contact process of complex network models[46], its role in inducing GP in brain networks remains unexamined. Consequently, how SC modularity and regional heterogeneity synergistically generate a biologically viable GP of brain networks that can accommodate individual variability into the critical regime remains unknown. Functionally, there is still no established mapping between an individual's position within the GP-derived biological plausible parameter range and their distinct functional repertoire, cognitive performance profile, or pathological vulnerability. These limitations perpetuate a fundamental question: Is GP merely a passive byproduct of brain network heterogeneity, or an evolved adaptive mechanism optimizing robustness-flexibility trade-offs for diverse cognitive demands of individual brains?

In this work, we addressed the aforementioned limitations by demonstrating that a synergistic interaction between SC modularity and regional heterogeneity generates an extended critical regime known as GP. This regime is uniquely characterized by the broad range in global excitability that varies across individuals and an embedded optimal trade-off point balancing global and local information transmission. These unique features of GP enable brain networks to simultaneously achieve robustness and flexibility in functional organization while accommodating individual variability in dynamics and function. Through analysis of resting-state functional magnetic imaging (rs-fMRI) data from the Human Connectome Project (HCP, n=990), we characterized the individual-specific synchronization patterns by demonstrating an inverted U-shaped dependence of synchronization entropy (SE) on mean synchronization (MS) both at global and local scales. We developed a whole-brain model with calibrated regional heterogeneity to show that individual-specific synchronization patterns can be primarily replicated by varying the global excitability of the model. Importantly, we provided evidence for the existence of GP in biological brains, as demonstrated by semi-independent asynchronous-synchronous transition among modules and critical dynamics with individual-specific critical exponents dependent on global synchronization level (excitability). Furthermore, we showed that the unique features of the GP in biological brains, i.e., the broad range in global excitability and a built-in optimal trade-off point balancing global and local information transmission, arise from the synergistic interactions between SC modularity and regional heterogeneity. This synergy endows brain networks with maximal robustness to variations in global excitability through stable critical dynamics, and flexible modulation of global and local information transmission. Consequently, the GP allows individual brains to operate at distinct global excitability levels, thereby supporting unique cognitive performance profile through flexible FC configurations that balance network segregation and integration.

**Results**

We first parcellated the whole brain into seven subsystems and characterized individual brain dynamics using individual-specific synchronization patterns that combine both global and local measures. We identified a semi-independent asynchronous-synchronous transition in brain networks, which implies a spatial asynchronous phase transition governed by GP. Next, we built a regionally heterogeneous whole-brain model with hierarchically calibrated local excitability to demonstrate that regional heterogeneity could induce GP in brain networks, and that individual brains could be mapped to a distinct position within the GP regime. Then, we used the whole-brain model to demonstrate that the synergistic interaction between SC modularity and regional heterogeneity in biological brains results in a GP with unique features that may empower brain networks to reconcile robustness and flexibility. Finally, we demonstrated that this unique feature of GP enables the individual brain to exhibit personalized functional and cognitive performance profiles by operating at distinct positions within or outside the GP regime.

**Individual variability in synchronization patterns and GP evidenced by semi-independent asynchronous-synchronous transition in empirical brain networks**

The brain was parcellated into seven functional modules (Fig. 1, top panel inside the dashed grey circle)

and further subdivided into 210 cortical and 36 subcortical regions of interest (ROIs; Fig. 1, brain maps on the dashed grey circle; see also Tab. S1 in Supplementary Information (SI) for detailed information on these ROIs and their anatomical regions), following the hierarchical tree organization of brain structures outlined by the Human Brainnetome (BN-246) Atlas[47]. Consequently, the derived brain networks encompass three hierarchical levels: whole brain, modules, and ROIs. Subsequently, blood-oxygen-level-dependent (BOLD) signals from the 246 ROIs, along with the voxel-level signals within each ROI, were extracted for analysis.

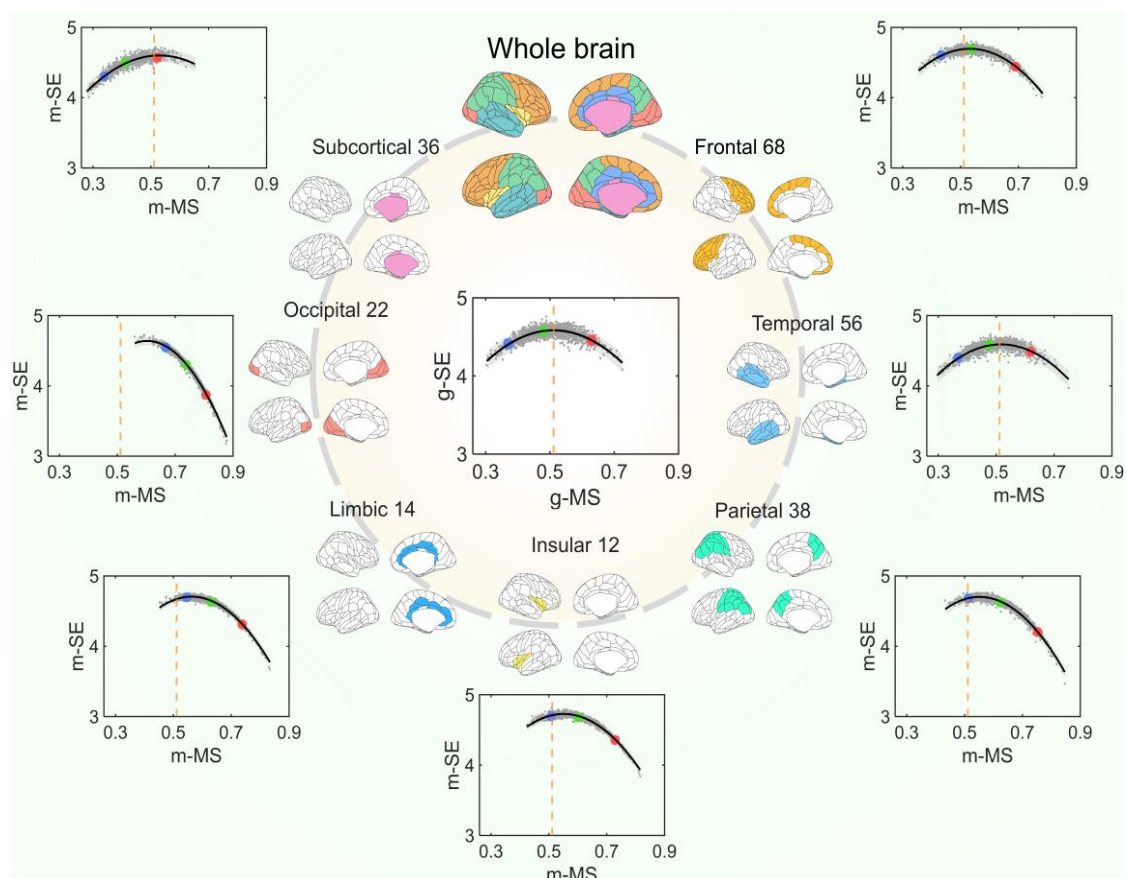

**Fig. 1 Global and local synchronization patterns across individual brain networks and the semi-independent asynchronous-synchronous (A-S) transition.** On the grey dashed circle: Top: the distribution of 7 modules parcellated according to the BN246 atlas, each denoted with a distinct color: frontal lobe (orange), temporal lobe (sky blue), parietal lobe (bluish green), insular lobe (yellow), limbic lobe (blue), occipital lobe (vermillion), and subcortical nuclei (reddish purple). The number and map of ROIs of each module are also shown respectively on the circle. Middle inside the grey dashed circles: The A-S transition curve for the whole brains with grey dots indicating individual global mean synchronization (g-MS) and global synchronization entropy (g-SE) values for each participant. Outside the grey dashed circle: The A-S transition curve for each module. In all the A-S plots, black lines represent quadratic fit, and grey shade areas represent the 95% confidence interval. Larger blue, green, and red filled circles denote average g/m-MS and g/m-SE values for three subgroups (n=30 each). The vertical orange lines indicate the tipping point of the A-S plot for the whole-brain network.

We first characterized the individual-specific patterns of global and local brain dynamics using two complementary measures of the extracted BOLD signals: mean synchronization (MS) to quantify average phase synchrony, and synchronization entropy (SE) to assess the diversity and complexity of synchrony over time. At the whole brain level, we performed the Hilbert transform on the BOLD signals of 246 ROIs for each participant to obtain their instantaneous phase. We then calculated the instantaneous Kuramoto order parameter to derive the global mean synchronization (g-MS) and global synchronization entropy (g-SE) for each participant (see Methods). As illustrated in the bottom panel inside the dashed grey circle of Fig. 1, the values of g-MS and g-SE across participants exhibited an inverted-U relationship on the g-MS *vs.* g-SE plot, which could be well-fitted with a quadratic function ($R^2 = 0.62, p < 10^{-10}$). This inverted-U curve reflects a considerable individual variability in global brain dynamics pattern, with a global A-S transition occurs with the moderate value of g-MS ($\approx 0.5$). Thus, this analysis established a relationship between individual difference and a dynamical transition of brain states with respect to the global synchronization level. At the module level, we calculated the module mean synchronization (m-MS) and module synchronization entropy (m-SE) for each of the seven modules per participant using the BOLD signals from the constituent ROIs of the modules. The resulting m-MS *vs.* m-SE plots for the seven modules are arranged clockwise outside the dashed grey circles in Fig. 1. Notably, quadratic functions provide good fits for these relationships (clockwise order: $R^2$= 0.84, 0.66, 0.96, 0.90, 0.97, 0.99, 0.83; $p < 10^{-10}$ in all cases). Taken together, these results demonstrate substantial individual variability in both global and local brain dynamics.

Theoretically, GP emerges in heterogeneous systems during spatial asynchronous transition from disordered to ordered phase, leading to the coexistence of order and disorder around the critical point[48, 49]. We therefore examined spatial heterogeneity in A-S transitions among modules for the evidence of GP in brain networks. To this end, we selected three participant subgroups (n=30 each) stratified by g-MS values (subgroup-averaged: 0.37, 0.48, 0.63; denoted by blue, green, and red filled circles in Fig. 1 g-MS vs. g-SE plots). For these subgroups, we computed subgroup-mean m-MS and m-SE values and plotted them in corresponding module-level panels of Fig. 1. We observed that as the whole-brain networks transition from asynchronous (blue dots) to synchronous state (red dots), passing through the transition point (green dots), though all modules exhibits a significant increase in their m-MS, modular transitions are semi-independent: Frontal and temporal modules transition in parallel with the brain's overall shift; Parietal, insula, limbic, and occipital modules progress from the transition point toward synchrony; Subcortical nuclei shift from asynchrony toward the transition point. Therefore, this empirically observed semi-independent A-S transitions across functional modules provides compelling evidence for the existence of GP in the human brain. Furthermore, the co-variation of global and modular dynamics suggests that the empirically observed A-S transitions across scales in brain networks are primarily driven by a single, system-wide parameter. In the following subsection, we developed a region-heterogeneous whole-brain model that formalizes global excitability as this key parameter and successfully replicated the above empirical observations. We also noted that while the above characterization of individual brain network dynamics is not entirely precise, as evidenced by residual variability in the m-MS of subsystems for a given g-MS level (Fig. S1), which potentially enable high-dimensional individual differences, this hierarchical framework nonetheless effectively captures the essential features of the GP of brain network, representing a significant advance over previous single-scale approaches[35].

**Moderate extent of regional heterogeneity induces Griffiths phase in the whole-brain model**

While rs-fMRI evidence supports the existence of GP in biological brain networks, does regional heterogeneity, in addition to that of SC, also contribute to its emergence? To address this gap in previous studies, we developed a regionally heterogeneous whole-brain model by extending the Haimovici et al. framework[50] with incorporated regional heterogeneity (see Fig. S2 for the illustration and Methods for of the modeling approach). Our model comprises 246 nodes (parcellated ROIs) with dynamics governed by discrete three-state Greenberg-Hastings cellular automata, interconnected via a SC matrix (246×246) derived from HCP DTI-based axonal fiber counts. To incorporate regional heterogeneity, we assume that excitability at nodal, modular, and global scale contributes additively to a node's effective excitability (Eq. (8) in Methods). Acknowledging that neural excitability scales proportionally with neural synchronization dynamics both globally and locally[37, 45, 51], we calibrated excitation threshold for each module or ROIs using participant-averaged MS derived from BOLD signals in the modules or ROIs (Eqs. (9) and (10)). The global baseline threshold $b$ was treated as a free parameter to model interindividual variability of brain network in global excitability. For simplicity, the extent of heterogeneity across modules ($K^M$) and ROIs within $J$-th module ($K_J^R$, $J$=1, 2, …, 7) was controlled by a single parameter $K$ (i. e., $K^M = K_J^R \equiv K$).

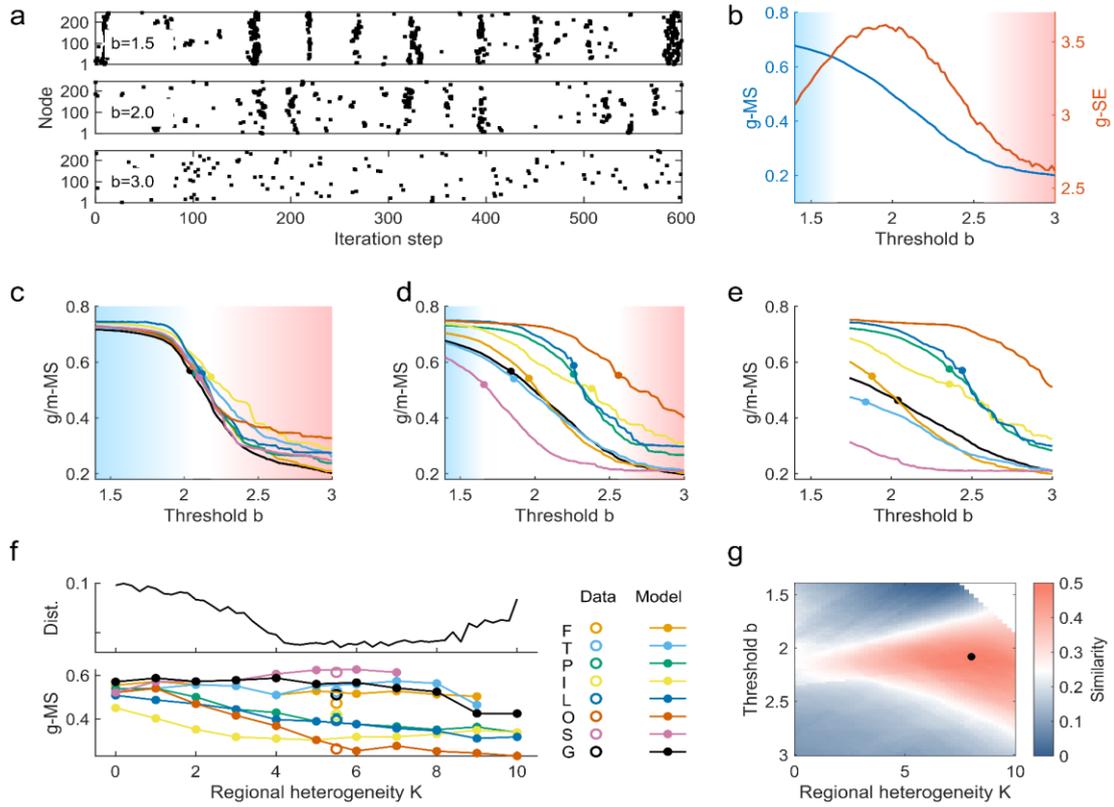

**Fig. 2 The whole-brain model with moderate regional heterogeneity replicated the semi-independent A-S transition among modules and most realistic FC patterns. a:** Raster plots of the excitation events of the models with different baseline thresholds $b$. **b:** The dependence of g-MS (blue line) and g-SE (orange line) among simulated 246 BOLD signals on the baseline threshold $b$ of the heterogeneous whole-brain model ($K$=6). **c-e:** The dependence of the g-MS (black line) and m-MS (colored lines) for each module on the baseline threshold $b$ for different $K$. **c:** $K$=0; **d:** $K$= 6; **e:** $K$= 8.6. The filled circles on the lines mark the transition points where the corresponding values of g/m-SE are maximized. The white area between red and blue shadow roughly demonstrated the extent of dispersion among the transition points of modules. **f:** (Bottom panel) The dispersed pattern of the global (G) and modular transition points (F=frontal, T=temporal, P=parietal, I=insular, L=limbic, O=occipital, and S=subcortical) predicted by the model with different $K$. The open circles between $K$=5 and $K$=6 mark the global and modular transition points observed from the fMRI data (see Fig. 1). The top panel displays the distance between the transition points in real data and model predictions with different $K$. **g:** The similarity between group-averaged empirical and simulated FC matrix as a function of $b$ and $K$. The filled back circle marks the position for the highest similarity. The white region denotes invalid parameter combinations (leading to negative threshold in model) where simulations failed.

Through varying the global baseline threshold $b$, the constructed whole-brain model with moderate heterogeneity $K$ successfully replicates the above empirically observed interindividual variability in brain network dynamics. As shown in Fig. 2**a**, decreasing $b$ (which increases global excitability) drives a disorder-order transition in excitation patterns of the model. Concurrently, an A-S transition emerges where g-MS for simulated BOLD signals increases monotonically (Fig. 2**b**, blue line) while g-SE peaks at the moderate $b$ values (Fig. 2**b**, brown line), thereby replicating the empirically observed inverted U-shaped relationship between g-SE vs. g-MS from Fig. 1. Crucially, comparing low- (e.g., $K = 0$; Fig. 2**c**) and high- heterogeneity models ($K = 8.6$; Fig. 2**e**), the moderate-heterogeneity model ($K \approx 6$; Fig. 2**d**) best replicates the brain's semi-independent A-S transition. It avoids both the tightly clustered (at small $K$) and excessively dispersed (at large $K$) transition points of the modules (colored dots in Fig. 2**c-e**, determined by peaking point of m-SE *vs.* m-MS curses of these modules). Specifically, the occipital lobe initiates the transition as $b$ decreases, followed by the insular, parietal, and limbic lobes; the frontal and temporal lobes synchronize with the global transition, while subcortical nuclei exhibit a lag. Furthermore, the model with moderate heterogeneity $K$ optimally reproduce empirically observed dispersion in global/modular transition points (Fig. 2**f**, all transition points are indicated by their corresponding g-MS values; in top panel the similarity between simulated and empirical dispersion is quantified with Euclidean distance between simulated and empirical g-MS values of these transition points). Additionally, the similarity between simulated and empirical FC matrix (see Methods) is also maximized with moderate value of $K$ (Fig. 2**g**). It is noted that excessive heterogeneity $K$ values produce invalid negative excitation thresholds in highly synchronized modules/ROIs—particularly when global baseline threshold $b$ is low. This constraint in the model causes the absent data in the upper-right corner of $K$-$b$ plot (Fig. 2**g**, white region).

We analyzed the avalanche statistics (cascades of excitation events and their probability distribution for size $S$ and duration $L$ as illustrated in Fig.3**a**) in the whole-brain model. Figure 3**b** shows that in the region-homogeneous model ($K = 0$), low global excitability (high $b$) produces subcritical dynamics (avalanche-size distributions skewed toward small events), whereas high global excitability (low $b$) results in supercritical dynamics (high frequency of large avalanches), and critical dynamics with power-law distributions ($P(S) \sim S^{-\alpha}$) emerge only at intermediate $b$. Conversely, the region-heterogeneous model exhibits critical dynamics with power-law avalanche-size distribution across a broader $b$ range (exemplified by $K = 6$ in Fig. 3**c**). We identified critical dynamics using the mean-squared deviation (MSD) between empirical avalanche-size distributions and their best-fit power-law function (exponents estimated from data, see Methods), adopting MSD $< 2 \times 10^{-5}$ as an *ad hoc* criterion (Fig. 3**d**). As shown in Fig. 3**e**, low heterogeneity (e.g., $K = 0$) confines power-law-distributed avalanches to a narrow range of $b$. Increasing regional heterogeneity expands the critical regime primarily into the subcritical domain, establishing a GP that bridges the subcritical and supercritical regimes. In addition to the size exponent $\alpha$, we also estimated the duration exponent $\tau$ from the distribution $P(L) \sim L^{-\tau}$, and a third exponent $\gamma$ from the average avalanche size $< S(L) >$ for a given duration $L$, where $< S(L) >= L^{\gamma}$. Within the GP regime, both $\alpha$ and $\tau$ decrease as baseline threshold $b$ increases yet maintain the scaling relation $\frac{\tau-1}{\alpha-1} = \gamma$ across varying extent of heterogeneity (Fig. 3**f**).

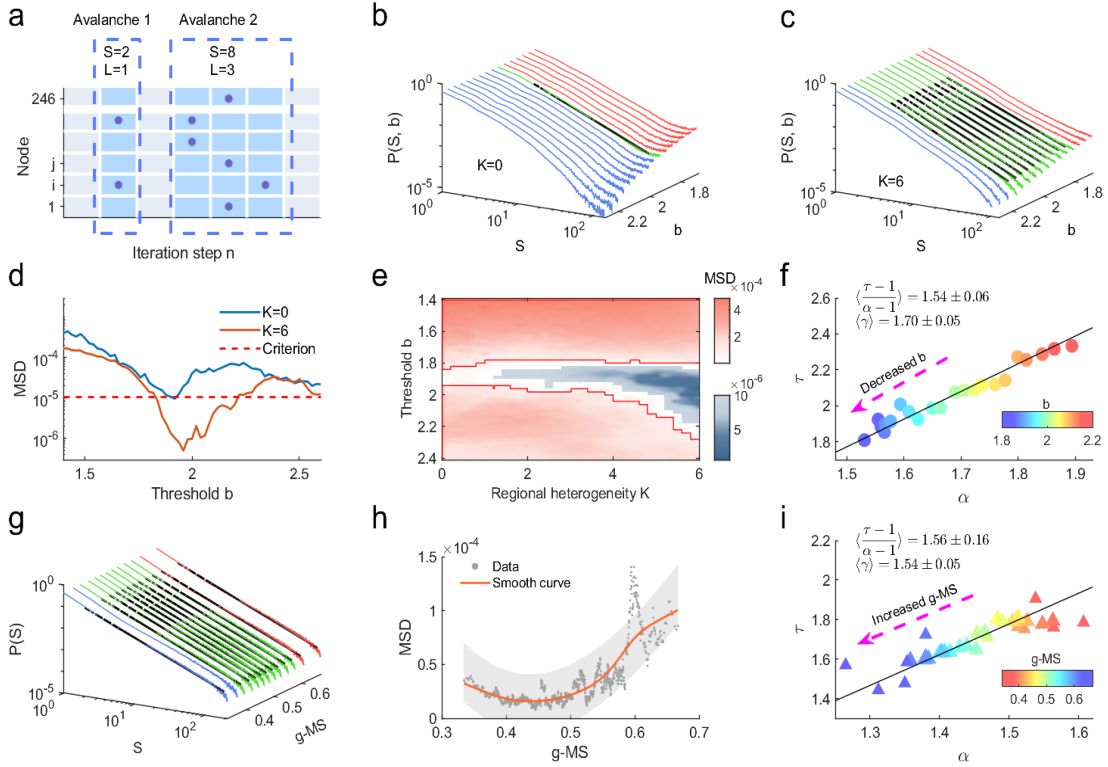

**Fig. 3 The stretched critical dynamics and Griffiths phase (GP) in the brain networks within which the scaling relation holds for individual-specific critical exponents. a**: Detecting the avalanches from the spatiotemporal excitation events in the brain networks. **b**: Probability distribution of avalanche sizes $S$ for different baseline threshold $b$ on the whole-brain models with $K$=0. **c**: The same as in **b** but for models with $K$=6. **d**: The mean-squared deviation (MSD) of the size distribution from the best power-law fitting as a function of $b$ for models with $K$=0 and $K$=6. We chose a value (horizontal dashed line, MSD=$2 \times 10^{-5}$) that is 2 times higher than the minimal MSD value of $K$=0 as the criterion for criticality. **e**: The GP regime (blue region) and its boundaries (red lines, MSD=$2 \times 10^{-5}$) defined by the dependence of MSD on both $b$ and $K$. **f**: The scaling relation obtained from the models within GP. The color of circles encodes the value of $b$, as indicated with color bar. **g**: The avalanche size distribution estimated from BOLD signals with a predefined binarizing threshold 2.0 SD (standard deviation) for different group-mean g-MS values. **h**: The MSD as a function of g-MS for avalanche size distributions estimated from BOLD signals in (**g**). The orange line is a cubic smoothing spline fit. **i**: The scaling relation calculated from the BOLD signals with the same predefined binarizing threshold 2.0 SD. In **g-i**, individuals were ranked by their g-MS values and grouped into subgroups of 30 participants with comparable g-MS values to calculate avalanche distributions and critical exponents for each group. Therefore, each pentagram in **i** denotes one subgroup, with subgroup-mean g-MS value indicated by the color bar.

Based on the above findings with the whole-brain model, we next investigated whether an individual's global brain state could be mapped to a distinct position within the GP regime. This was assessed by examining the g-MS dependence of critical exponents derived from BOLD signal avalanche statistics. To mitigate estimation errors in these statistics due to limited data length in individual brain scans, we ranked the subjects according to g-MS value and divided the subjects at hand into 33 subgroups, each containing 30 participants with similar g-MS values. Considering the monotonic relationship between $b$ and g-MS (Fig. 2**b**), we expected to observe a g-MS-dependent variation in power-law exponents $(\alpha, \tau)$ across individual in rs-fMRI data, with these variations adhering to the scaling relation. We then computed the avalanche size distribution for each subgroup from aggregated neural avalanche data across all participants within the subgroup (Fig. 3**g**). We observed that the estimated MSD values for avalanche size distribution are minimal for subgroups with moderate g-MS values ($\approx 0.45$, Fig. 3**h**). As expected, the empirically derived critical exponents fall along the line predicted by the scaling relation, with $\alpha$ and $\tau$ decreasing as the g-MS increases (Fig. 3**i**). It is noted that the empirically estimated critical exponents are also affected by the threshold used to binarize BOLD signals, with higher thresholds resulting in larger estimated critical exponents (Fig. S3). Specifically, a higher threshold would splinter large avalanches into smaller ones. With fewer large avalanches to define the tail of the distribution, the slope becomes steeper, leading to an increase in the effective critical exponent. Importantly, the conclusions that critical exponents fall along the line predicted by the scaling relation is robust irrespective of the dependence of estimated critical exponents on the binarization thresholds (Fig. S3).

**Unique feature of GP in brain networks: optimal point in broad GP range for the balanced global/local information transmission**

Having established regional heterogeneity as a significant driver of GP, we next employed the whole-brain model to investigate its synergy with SC modularity in regulating information transmission and modulating GP. We quantified information transmission on brain networks using avalanche propagation matrices (APM) across the 246 ROIs, which measure the probability that excitation in one node triggers excitation in another (top panels, Fig. 4**a** and **b**; for better visualization, the APMs are presented in their coarse-grained form across modules, see Methods for details). According to Eqs. (9) and (10), increasing $K$ diversifies excitation thresholds across modules and ROIs, thereby amplifying disparities in intra-modular information flow between distinct brain modules (Fig. 4**a**, second-row panels). Consequently, the inter-module variation in information flow (indicated by arrows marking slopes in second-row panels of Fig. 4**a**) increases with $K$ (blue line in bottom panel of Fig. 4**a**). We also adjusted the SC modularity ($Q$) of the original SC matrix by strategically rewiring the randomly chosen connections to increase or decrease inter-module connections while preserving the total input strength of each nodes (see Methods). We noted that the homeostatic principle incorporated into the whole-brain models ensures that manipulating the SC matrix preserves the average SC connection strength, allowing models built on these adjusted SCs to remain comparable. We observed that as $Q$ increases, the APMs exhibit heightened modularity (Fig. 4**b**, second-row panels), indicating a stronger preference for intra-modular over inter-module propagation (blue line, bottom panel of Fig. 4**b**).

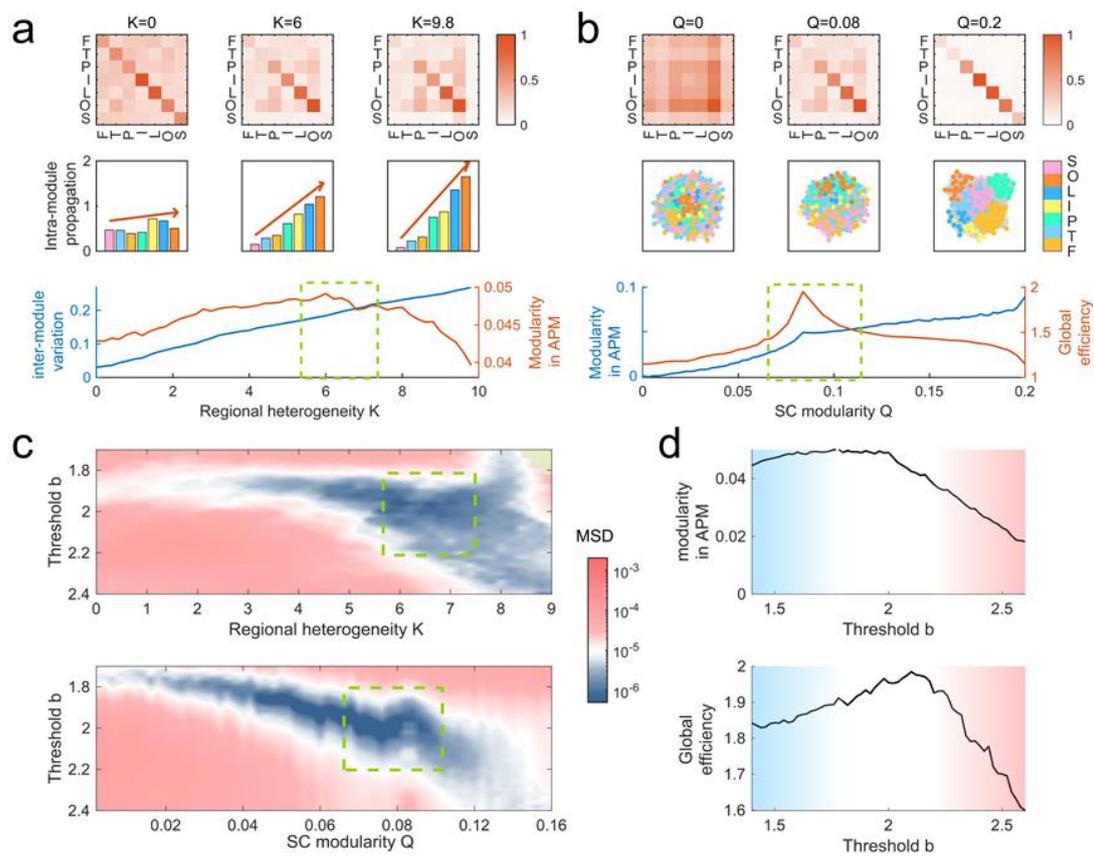

**Fig. 4 The precise interplay between SC modularity and regional heterogeneity yields a wide GP regime containing an optimal point that balances local and global information flow. a**: The regional heterogeneity dependent avalanche propagation in the whole-brain models. Top: the avalanche propagation matrices among the seven modules for different extent of heterogeneity. Middle: the distribution of intra-modular propagation for the seven modules. The arrows indicate the slope of linear fitting of these distributions, representing inter-module variability in information flow. Bottom: the dependence of inter-module variability (slope in the middle panel), as well as modularity in avalanche propagation matrix as a function of regional heterogeneity $K$. The green dashed box marks the GP in the biological plausible range for $K$. In the simulations, $b$=2.0 and $Q$=0.08. **b**: The SC modularity dependent avalanche propagation in the whole-brain models. Top: the avalanche propagation matrices for different extent of modularity $Q$ in the SC network. Middle: A low dimensional visual representation of the modularity in

avalanche propagation matrices for different extent of SC modularity $Q$. Each dot represents each of the 246 nodes, while the color code and spatialization highlight the network modularity. Bottom: The dependence of the modularity and global efficiency of the avalanche propagation matrices as a function of SC modularity $Q$. The green dashed box marks the GP in a range of $Q$ around the real brain SC network. In the simulations, $b$=2.0 and $K$=6.0. **c**: The GP regime (blue region) in the parameter space of $(b, K)$ (top panel) and $(b, Q)$ (bottom panel**)**. The green dashed boxes indicate the critical regimes ($b \sim 1.8$-$2.2$) corresponding to the biologically plausible ranges of $K$ and $Q$ as in **a** and **b**. The light green region in the upper panel denotes invalid parameter combinations (see Fig. 2**g** for details). **d**: The dependence of modularity (top) and global efficiency (bottom) of the avalanche propagation matrices on $b$. The white area between red and blue shadow corresponds to the critical regime ($b \sim 1.8$-$2.2$) indicated by the green dashed boxes in **c**.

Although increases in SC modularity $Q$ boost localized information transmission, the default SC architecture in the real brain ($Q$=0.08) inherently favors global communication[52], largely due to its economical small-world topology[53, 54]. This is evidenced by APMs achieving peak global efficiency at the default SC modularity value of the real brain ($Q$=0.08; orange line in bottom panel of Fig. 4**b**). This finding prompts a key question: how do brain networks balance the competing demands of localized processing and global integration of information, a prerequisite for flexible cognition and adaptive behavior? Our analysis revealed that modularity in the APMs of the region-homogeneous model ($K$=0) stemmed exclusively from the intrinsic modularity of the SCs. Increasing $K$ enhances modularity in the APMs by promoting module independence through regional excitability differences. However, excessive regional heterogeneity (high $K$) transformed low-threshold regions (e.g., the occipital lobe) into dominant hubs by persistently activating their neighboring ROIs, thereby disrupting modular organization and reducing overall modularity in the APMs. Consequently, the APMs exhibited peak modularity at a range of intermediate heterogeneity (green dashed box around $K \approx 6$), shown by the orange line in bottom panel of Fig. 4**a** ($b$=2.0). This range of $K$ with peak modularity aligns with the range reproducing the empirical semi-independent A-S transition and FC of the real brain (Fig. 2**f**, **g**). This indicates that the extent of regional heterogeneity in real brains also represents a tuned optimum for facilitating localized information processing, and that the precise coupling of SC modularity and regional heterogeneity underpins the brain's ability to resolve competing demands between localized processing and global integration.

How does the interplay between SC modularity and regional heterogeneity influence the GP regime? We found that the synergy of structural modularity and regional heterogeneity produces a GP that can sustain stable critical dynamics across a broad range of global excitability (green dashed box in Fig. 4**c**).

Importantly, both modularity and global efficiency of the APMs peak at an optimal excitability level in this wide GP regime ($b \approx 2.0$, Fig. 4**d**), suggesting that this point also represents a balance between the competing demands for segregated local processing and integrated global communication. Taking together, these results suggest that the GP in brain networks is uniquely characterized by both broad range of global excitability that preserves critical dynamics and an embedded optimal point that balances local and global information transmission. We hypothesized that by operating around strategic points within the GP regime, individual brain networks can be flexibly configured to prefer global, local, or balanced information processing, thereby favoring specific functions. Concurrently, sustained critical dynamics ensure stable performance of distinct functions across all operational modes. We thus argued that this unique feature of GP in brains provides a dynamical foundation for brain networks to embrace both robustness and flexibility across individuals.

**GP accommodates individual difference and predict diverse cognitive abilities**

Using the GP boundaries at $K$=6 (Fig. 3**e**) and the monotonic relationship between $b$ and g-MS (Fig. 2**b**) obtained from the whole-brain model, we estimated the g-MS range for GP in real brains as $0.38 <$ g-MS $< 0.58$. We then categorized the 990 participants into three groups based on their g-MS values: subcritical (g-MS $< 0.38$), supercritical (g-MS $> 0.58$), and the GP group ($0.38 \leq$ g-MS $\leq 0.58$; ~78% of participants, Fig. 5**a**). We further subdivided the GP group into three equal-size subgroups (GP1: $0.38 <$ g-MS $< 0.46$; GP2: $0.46 <$ g-MS $< 0.52$; GP3: $0.52 <$ g-MS $\leq 0.58$).

From each participant's cortical FC network, we measured their FC diversity[9] charactering the broad range of pairwise correlation, along with eigenmode-based FC network segregation ($H_{Se}$), integration ($H_{In}$), their balance ($H_B$) and temporal flexibility in the transition between segregated and integrated states [11] (see Methods). We also estimated latent factors of general ability and three domain-specific cognitive abilities (crystallized intelligence, processing speed and memory) from nine specific task performance indicators using structural equation modeling (SEM). Subsequently, within a multiple-group modeling framework[7] (see Methods), we examined the predictability of the GP's unique feature for these cognitive abilities across individuals. We found that participants with a moderate g-MS value (~0.5) maintained the most balanced state between FC segregation and integration (Fig. 5**b**, top two panels) and exhibited the highest levels of flexibility (Fig. 5**b**, third panel) and diversity (Fig. 5**b**, bottom panel; see Methods) in their FCs. These participants also demonstrated peak performance in memory tasks (Fig. 5**c**, bottom-right panels). These results suggest the existence of an optimal working point in brain networks that balances the trade-off between functional segregation and integration, thereby maximizing specific cognitive functions like memory. Meanwhile, while supercriticality with stronger FC integration is associated with better general cognitive ability across the g-MS range, and subcriticality with stronger segregation fosters crystallized intelligence and processing speed, both general cognitive ability and crystallized intelligence exhibits nonlinear dependence on g-MS, as shown by the superior fit of a cubic function over a linear one (cubic $R^2 = 0.89$ and $0.76$ vs. linear $R^2 = 0.70$ and $0.53$; Fig. 5**c**, top panels). Thus, the observation of maximized memory performance at an optimal g-MS configuration, combined with the relatively stable performance of general cognitive ability and crystallized intelligence scores across the GP regime, suggests a correlation between the unique features of GP and individual robustness and variability in cognitive performance profiles, implying the potential role for GP-mediated brain networks in meeting personalized cognitive demands.

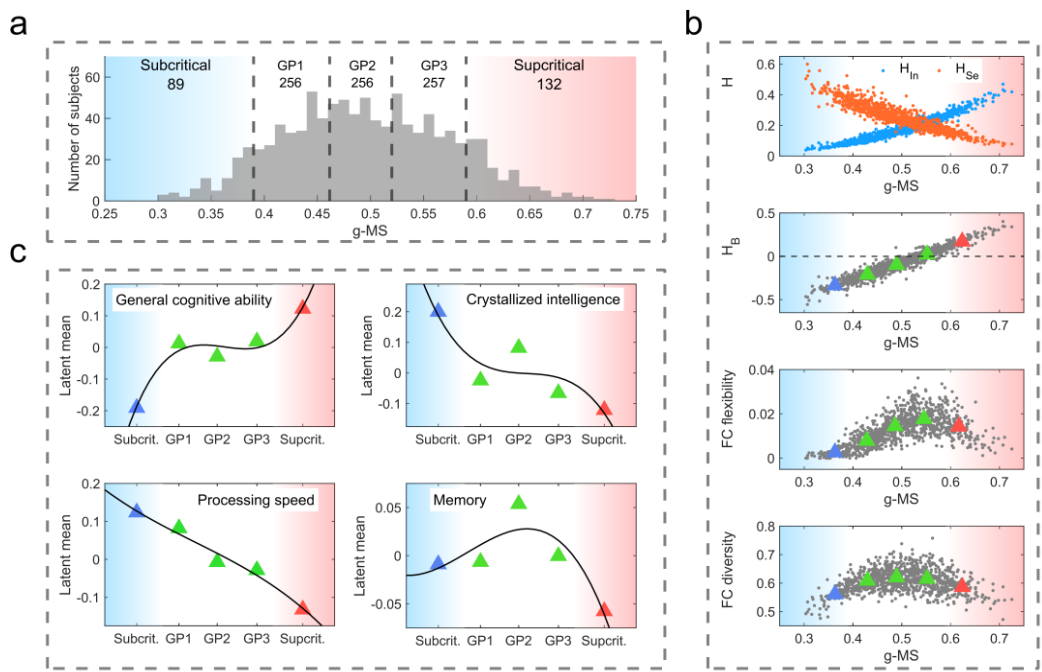

**Fig. 5: GP provide the brain with the capacity to accommodate robust individual differences in cognitive capacities. a**: The classification of 990 subjects into three regimes: the subcritical, GP, and supercritical regimes. The margins of the red and blue shadows indicate the lower and upper boundaries of g-MS for the GP regime. The dashed lines further divide the subjects in the GP regime into three subgroups. **b**: The dependence of individual cortical FC network properties on g-MS values derived from their regional BOLD signals. Panels show, from top to bottom: segregation ($H_{Se}$), integration ($H_{In}$), balance ($H_B$), flexibility, and diversity. **c**: The SEM estimated general cognitive ability (left-top), crystallized intelligence (right-top), processing speed (left-bottom), and memory (right-bottom) for groups of participants from the subcritical, supercritical regimes, as well as three subgroups in the GP regime. Black lines in **c** represent cubic fit of the data ($R^2 = 0.89$ for general cognitive ability, $0.76$ for crystallized

intelligence, 0.96 for processing speed, and 0.82 for memory). In contrast, the coefficient of determination ($R^2$) for the linear fit was 0.70, 0.53, 0.91, and 0.17, respectively.

## Discussion

This work establishes that GP is a comprehensive brain criticality framework to accommodate individual difference in biological brain dynamics and enable brain networks to reconcile robustness and flexibility to meet diverse cognitive demands across individuals. Through analysis of large-scale rs-fMRI data, we characterized individual differences in brain network dynamics by identifying each participant's unique position along a global/local A-S transition curve and their individual-specific critical exponents. Using a whole-brain model incorporating biologically plausible regional heterogeneity and empirical SC, we demonstrated that systematically varying global excitability induces an emergent GP which primarily accounts for the empirically observed individual differences in brain network dynamics. Furthermore, we revealed that the synergy between biologically plausible SC modularity and regional heterogeneity produces an optimal GP spanning a broad range of global excitability and containing an optimal point balancing global and local information transmission. We proposed that this unique feature of the GP enables the brain network to reconcile robustness and flexibility, a core principle in systems biology, thereby supporting diverse cognitive profiles for individual human. We empirically supported this hypothesis by demonstrating a correlation between this unique GP feature and individual cognitive performance profiles. Therefore, by addressing two critical gaps in prior research, the significant regional heterogeneity and individual differences, our work greatly advances the GP framework in the understanding of brain criticality in both mechanistic and functional perspectives. This improved framework of brain criticality offers profound implications for network and cognitive neuroscience across healthy and pathological states, while advancing physics-grounded brain network modeling.

The classical brain criticality theory provides a methodological framework to understand the brain's complex spatiotemporal dynamics[14, 55]. However, the brain is a quintessential nonequilibrium system: it dynamically interacts with its environment, continuously dissipates energy through metabolism, and self-organizes into structural and functional hierarchies—ranging from molecular interactions to large-scale neural networks[56, 57, 58] which vary significantly across individuals. However, the dominant criticality framework, often grounded in simplified models like the sandpile, branching process, or Ising model, fails to fully account for the intricate, inherently non-equilibrium dynamics characteristic of biological brains[59]. The empirically observed variation in critical exponents, which consistently adhere to scaling relations, serves as a touchstone for evaluating evolving theories addressing this gap[28]. Several solutions has been proposed[27]: Subsampling theory attributes variations in critical exponents to measurement artifacts[60]. Quasicriticality theory recognizes that living neural systems continuously process external inputs, predicting a dynamic departure from the critical point, manifested through stimulus-dependent scaling exponents which still adhere to the scaling relation[28, 61]. Complementarily, the earlier GP theory posits that modular architecture in the SCs leads to an extended critical regime[31], within which the critical exponents depend on the control parameter but still adhere to scaling relation[34]. In this work, by linking the control parameter of a whole-brain model to global excitability across individuals and integrating both structural and nodal heterogeneity in brain networks, we provided a more comprehensive and biological plausible picture of brain critical dynamics than prior studies. Crucially, this updated GP framework explains the empirically observed dependence of critical exponents on individual synchronization level (g-MS) as arising from individual variations in global excitability. This framework also holds potential to explain the previously reported variation in critical exponents across different contexts (e.g., across species, within species over time, and across stimuli) by incorporating variations in structural modularity and regional heterogeneity across species[62], as well as endogenous and exogenous input-dependent dynamic modulation of global excitability in individual brains[28, 63, 64]. Theoretically, the renormalization group (RG) serves as a powerful framework for understanding criticality, classifying diverse systems across physical and complex domains into universality classes based on their invariance under RG transformations[65]. In neuroscience, it holds significant promise for explaining scale-invariant, critical brain states and the connection between microscopic interactions and macroscopic network dynamics[66], thereby illuminating the brain's multi-scale organization[67]. Our work suggests a crucial improvement to the current approaches: the brain's profound heterogeneity, which is manifested in regional diversity and connectome variations, results in non-universal critical dynamics. We contend that progress in the field necessitates enhanced RG frameworks that explicitly incorporate this heterogeneity, moving beyond the assumption of a uniform critical point to explain the individualized critical landscapes that underlie cognitive function[68].

The classical brain criticality theory also offered a compelling framework for understanding how brain networks function[69, 70, 71, 72]. However, the paradigm of a single critical point faces significant challenges in explaining the empirically observed individual variability in brain dynamics and personalized cognitive profiles[16, 73, 74]. Systems biology reveals that this variability is not mere noise, but an inherent feature of biological networks, essential for reconciling system-wide robustness with flexibility[23, 24, 75]. While predictions derived from the classical single-critical-point model are increasingly at odds with this systems biology perspective, our work demonstrates that an updated GP framework successfully bridges this gap. The GP framework inherently accommodates individual variability and empowers brain networks to simultaneously embrace both robustness and flexibility. This aligns with the core evolutionary principle of biological systems elucidated by systems biology[23]. Notably, prior GP research has begun to address aspects of this gap. Early work indicated that GP extends the functional advantages associated with a critical point across a broad parameter regime[31]. While a more recent study specifically highlighted GP's capacity to account for individual variability in brain dynamics[35], it attributed the phenomenon solely to heterogeneity in SCs and explained the variability in synchronization levels by an individual's position along the GP, where healthy brain areas operate in a subcritical state and epileptogenic areas in a supercritical one. As we argued in the Introduction, a critical factor neglected in this and previous studies is the contribution of regional/nodal heterogeneity to the emergence of GP. Bridging this gap is essential to achieve a complete picture of the GP in brain networks and elucidate how it confers both robustness and flexibility to brain networks. In our work, by integrating regional heterogeneity into a simple but biologically plausible whole-brain model, we provided a clearer and more complete picture on how GP accommodates individual variability in brain dynamics through the interplay among three key biological determinants (global excitability, SC modularity, and regional heterogeneity). Specifically, the GP with the broad range of global excitability driven by biologically meaningful extent of SC modularity and regional heterogeneity implies that GP is not a passive byproduct of network heterogeneity. Instead, it represents an evolved, robust mechanism that allows for individual variability. Furthermore, we showed that by adjusting their position around an optimal point within GP, individual brains could flexibly negotiate a trade-off between local and global information transmission or between FC segregation and integration. Crucially, based on this mechanistic understanding of GP, we further found there is a correlation between this unique feature of GP and individual cognitive performance profiles, suggesting the vital role of GP playing in meeting the diverse personalized cognitive demands of biological brains. Therefore, our findings significantly advance this topic by demonstrating how the GP framework explicitly integrates the systems biology view of inherent individual variability as a fundamental mechanism for balancing robustness and flexibility in brain function.

By moving toward a more biologically plausible theory of brain criticality, the improved GP framework yields more accurate characterization of empirical brain dynamics. This advancement thus holds significant potential to broaden applications of criticality theory in multiple fields, particularly in brain disease research and digital twin brain technology. In the context of brain disease, it overcomes a key limitation of traditional single-point criticality models, which cannot account for disease-specific deviations from critical dynamics. The classical view holds that while criticality is a hallmark of healthy brains[55], and its pathological breakdown is interpreted as a deviation from an optimal dynamic point accompanied by a failure in supporting mechanisms such as excitation-inhibition (E-I) balance[14, 36]. However, this view presents a paradox: although criticality is indeed disrupted in various pathological states[13, 14], the brain also exhibits remarkable resilience to specific perturbations, such as stroke[76]—a phenomenon not adequately explained by classical criticality theories that rely on a single optimum point governed solely by E-I balance. The updated GP framework uniquely enables us to decode links between disease-specific brain network abnormalities (e.g., global excitability, SC modularity, and regional heterogeneity) and aberrant critical dynamics, and quantify resulting impairments in global and local information processing and robustness-flexibility balance, thereby establishing new mechanistic pathways for understanding criticality pathology in neurological disorders and providing a theoretical basis for future novel therapies[77, 78]. Similarly, the GP framework offers advantages for developing digital twin brains. Digital twin brains, as a cutting-edge focus in computational neuroscience, leverage multi-scale, multi-modal biological data to create personalized, dynamic simulations of biological brains, accelerating discoveries in neuroscience, therapeutics, and AI[79, 80]. However, neuronal/synaptic-level simulations demand prohibitive computational resources[81], while cost-effective generative network models often face performance limitations[82]. Our work demonstrates that incorporating realistic physics constraints (GP instead of a single critical point) enhances brain network model performance in capturing individual brain dynamics without increasing computational cost, underscoring the advantage of physics-grounded approaches.

In conclusion, by address two critical gaps on brain criticality, we combined rs-fMRI analysis and whole-brain modelling techniques to show that critical dynamics in biological brains are not only supported by global excitability as previous single-critical-point theory has assumed, but also shaped by structural modularity and regional heterogeneity in brain networks. The synergy among these three brain network factors results in an optimal GP with unique feature that enable the brain networks to maximally accommodate individual variability in brain dynamics and embrace both robustness and flexibility to meet diverse cognitive demands across individuals. This biologically plausible framework opens broad avenues for exploring implications of brain criticality in neural function and neurological pathogenesis, while promoting the development of computationally efficient physics-driven brain simulation techniques.

## Methods

### HCP dataset and preprocessing

*MRI data acquisition:* This work utilized data from the publicly available HCP 1200-subject release, which includes structural MRI, diffusion-weighted MRI, rs-fMRI, and behavioral measures from nearly 1200 subjects[83]. All HCP experimental procedures were conducted in accordance with ethical guidelines and were approved by the Washington University–University of Minnesota (WU-Minn) Institutional Review Board, with written informed consent obtained from all participants. Full details on the data acquisition protocol for HCP dataset have been published previously[84, 85]. For the current analysis, we included 990 subjects who underwent all the MRI scanning sessions, including four sessions of rs-fMRI and the diffusive MRI, and have completed all the neuropsychological testing.

*Brain parcellation:* Using the BN-246 atlas (http://atlas.brainnetome.org/bnatlas.html), we performed a hierarchical parcellation of the human brain into seven functional network modules: the frontal (F), temporal (T), parietal (P), occipital (O), and insular (I) lobes, along with limbic (L) areas and subcortical (S) structures. These modules were subsequently subdivided into 210 distinct cortical and 36 subcortical subregions (ROIs)[47]. The BN-246 atlas is a cross-validated, fine-grained atlas with information on both functional and anatomical connections, which not only confirmed some differentiation from early cytoarchitectonic mappings but also revealed many anatomical subdivisions that were not described previously.

*MRI data preprocessing*: The rs-fMRI data and DTI data preprocessed by the HCP through the well-established and -accepted Minimal Processing Pipeline (MPP)[84]. Furthermore, the rs-fMRI data was filtered by a fifth-order Butterworth bandpass filter with cutoff frequencies of 0.01 and 0.1 Hz. Afterwards，the BOLD signals from the 246 ROIs, along with the voxel signals within each ROI, were extracted for further analysis.

The DTI data of the first 100 selected subjects were reconstructed in the Montreal Neurological Institute (MNI) space using q-space diffeomorphic reconstruction (QSDR)[86], as implemented in DSI Studio (https://www.dsi-studio.labsolver.org) with the default parameters. A diffusion sampling length ratio of 1.25 was used. Restricted diffusion was quantified via restricted diffusion imaging [87], and a deterministic fiber tracking algorithm [88] generated one million fibers with whole-brain seeding. The angular threshold was randomly selected from 15° to 90°, and the step size from 0.1 to 3 voxels. Fiber trajectories were smoothed by averaging the propagation direction with a randomly selected percentage (0% – 95%) of the previous direction. Tracks shorter than 5 mm or longer than 300 mm were discarded.

**The empirical SC, FC, and simulated FC matrices**

*SC matrix:* For each subject, the SC matrix was computed by counting the number of streamlines connecting every pairwise combination of ROIs defined by BN-246 atlas. Then, subject-specific SC matrices were aggregated into a group-averaged SC matrix $W_{ij}$.

*FC matrix:* The empirical/simulated FC matrix was constructed by computing the absolute value of the Pearson correlation coefficient between the corresponding empirical/simulated BOLD signals of each pair of ROIs.

*Similarity measure*: The similarity between the matrices was quantified by reshaping their upper triangular elements into vectors and computing the Pearson correlation coefficient between these vectors.

*Manipulation the modularity of SC networks:* We manipulated the original SC matrix by randomly selecting a certain number of connections with a rewiring probability, breaking them, and redistributing their weights to other randomly selected target connections. If the target connections were chosen from the entire SC matrix, this manipulation would decrease the network modularity. Conversely, if the target connections were selected exclusively within the same module (one of the seven predefined modules), this manipulation would enhance the network modularity. The modularity is changed by varying the rewiring probability.

*The modularity of SC:* We quantified the modularity for both the original and manipulated SC matrices and avalanche propagation matrices in the perspective of weighted network with Newman's algorithm[89]:

$$Q = \frac{1}{2m} \sum_{ij} [A_{ij} - \frac{k_i k_j}{2m}] \delta_{c_i, c_j},$$ (1)

where $k_i$ is the degree of node $i$, $m$=60516 is the total number of edges (elements) in the network, and $A_{ij}$ is an element of the SC or avalanche propagation matrices. Here, $\delta_{ij}$ is the Kronecker delta symbol, and $c_i$ is the label of the modules to which node $i$ is assigned. The modules are the seven predefined functional modules in the BN-246 atlas.

**Identification of the A-S transition among modules**

For a given set of $m$ neural signals $F_j(t)$ ($j = 1, 2, m$), the Kuramoto order parameter could be calculated to quantify synchrony, alongside its Shannon entropy to assess variability in synchrony across time. The instantaneous phase trace $\theta_j(t)$ of signal $F_j(t)$ is obtained through the Hilbert transform $H[F_j(t)]$:

$$\theta_j(t) = arctan \frac{H[F_j(t)]}{F_j(t)}.$$ (2),

The Kuramoto order parameter $r(t)$ is then calculated as

$$r(t) = \frac{1}{m} \left| \sum_{j=1}^{m} e^{i\theta_j(t)} \right|.$$ (3)

Then, $\bar{r}$, the MS among these signals is defined as,

$$\bar{r} = \frac{1}{l} \sum_{k=1}^{l} r(t_k),$$ (4)

where $l$ is the total time points and $t_k$ is the $k$-th time point of the signals. Additionally, the SE of these signals is calculated from the probability distribution of $r(t)$ as

$$H(r) = - \sum_{i=1}^{n} p_i log_2 p_i,$$ (5)

where $p_i$ is the probability of $r(t)$ falling into the $i$-th bins when the [0, 1] interval is divided into $n$ bins. In this work, for the analyzed BOLD signals, $l$=1200 and $n$ was set to 30.

In this work, the whole-brain network synchrony and entropy, g-MS and g-SE, were calculated using BOLD signals from the 246 parcellated ROIs, where $m = 246$. For each module, the m-MS and m-SE were computed using the BOLD signals from the ROIs within that specific module, with $m$ representing the number of ROIs within each module. Specifically, $m$ =68, 56, 38, 12, 14, 22, and 36 for frontal, temporal, parietal, insular, limbic, occipital lobe, and subcortical nuclei, respectively. Furthermore, to calibrate the ROI-specific excitability, we computed the MS across voxelwise BOLD signals within each ROI. In this case, *m* represents the number of voxels in the given ROI, which ranges from 86 for the smallest ROI to 1534 for the largest, with a mean value of 571.9.

**The whole-brain model with regional heterogeneity**

*The original whole-brain model*: The whole-brain network model employed in this study was adapted from the framework developed by Haimovici et al.[50], which assumes identical intrinsic excitation thresholds across all ROIs. The nodal dynamics were governed by the GH cellular automaton, a three-state discrete reaction-diffusion model. At any time, each node could be in one of the three states: inactive (Q), excited (E), or refractory (R), and update their state at each iteration step according to the following rules (Fig. S2**d**):

$$
\begin{cases}
Q \rightarrow E & if \sum_j W_{ij}s_j(n) > T_i \text{ or with prob. } r_1 \\
E \rightarrow R & \text{with prob.1} \\
R \rightarrow Q & \text{after } n_{\text{delay}} \text{ and with prob. } r_2,
\end{cases}
\tag{6}
$$

where $s_j(n) \in \{0, 1\}$ represents the state of node $j$ at iteration step $n$. If node $j$ is in the "E" state, $s_j(n) = 1$; otherwise, $s_j(n) = 0$. In original model, the excitation threshold $T_i$ is uniform across all nodes *i*. If the conditions for state transition are not satisfied, the node retains its current state. $W_{ij}$ is symmetric and weighted SC matrix obtained from DTI/DSI scans of the white matter fiber tracts or the manipulated SC with rewiring. In this work, *i, j*=1, 2, …, 246.

*The homeostatic principle:* A previous study suggested that, at the whole-brain level, homeostatic plasticity mechanisms help maintain a balance between excitation and inhibition within the brain network. It also showed that incorporating this homeostatic principle into the model led to more realistic predictions of empirical FC pattern[90]. Therefore, following this study, we replaced the original SC matrix with the one that derived from the following normalization rule[90]:

$$
\widetilde{W}_{ij} = \frac{W_{ij}}{\sum_j W_{ij}}.
\tag{7}
$$

The normalized SC matrix $\widetilde{W}_{ij}$ was then scaled by a factor of 20 to match the model dynamics.

*Regional heterogeneity calibration:* We introduced node-specific thresholds into the original model to capture regional heterogeneity. Specifically, considering the hierarchical modular origination of the brain networks, we assume $T_i^J$, the excitation threshold of *i*-th ROIs in *J*-th module, is determined together by its intrinsic threshold $T_i^R$, the modular threshold $T_J^M$, and the global excitability $b$, *i.e.*,

$$T_i^J = T_i^R + T_J^M + b, \tag{8}$$

where $i = 1, \ldots, N_J$, and $N_J$ is the number of ROIs in the $J$-th module, and $J = \mathrm{F, T, P, O, I, L, S}$, corresponds to the seven modules defined by BN-246 atlas.

In line with established literature demonstrating that neural excitability scales proportionally with neural synchronization dynamics both globally and locally[37, 45, 51], we calibrated the above-defined excitation thresholds at multiple scales using BOLD-derived synchronization levels among neural activities. Specifically, we calibrated the excitation threshold $T_i^R$ for the *i*-th ROI in the *J*-th module with $< \overline{r_i^R} >$, the group-averaged MS among voxelwise BOLD signals within *i*-th ROI (Fig. S2**h** and bottom panel of Fig. S2**g**):

$$T_i^R = -K_J^R \cdot (< \overline{r_i^R} > - C_J^R), \tag{9}$$

where $\overline{r_i^R}$ represents the MS calculated from the voxelwise BOLD signals across all voxels in the *i*-th ROI, and $\langle \cdot \rangle$ denotes the average across subjects. The parameter $K_J^R$ controls the extent of heterogeneity at the ROI level in the *J*-th module. To ensure that the mean excitability across ROIs remains invariant to $K_J^R$, we demeaned $< \overline{r_i^R} >$ by subtracting $C_J^R$, where $C_J^R$ is the mean of $< \overline{r_i^R} >$ over all ROIs in the *J*-th module, i.e., $C_J^R = \frac{1}{N_J} \sum_{i=1}^{N_J} < \overline{r_i^R} >$. Similarly, we calibrated the excitation threshold $T_J^M$ for the *J*-th module with $< \overline{r_J^M} >$, the group-averaged m-MS among BOLD signals of the ROIs in the *J*-th module (Fig. S2**f** and top panel of Fig. S2**g**):

$$T_J^M = -K^M \cdot (< \overline{r_J^M} > - C^M), \tag{10}$$

where $\overline{r_J^M}$ is the m-MS of the *J*-th module, $\langle \cdot \rangle$ denotes the average across all subjects. The parameter $K^M$ controls the extent of heterogeneity at the module level, and $C^M = \frac{1}{7} \sum_{J=1}^{7} < \overline{r_J^M} >$, which is subtracted from $< \overline{r_J^M} >$ to ensure that the mean excitability across modules remains invariant under variations of $K^M$. For the baseline threshold $b$, rather than calibrating it with group-averaged g-MS, we designated it as a free control parameter to model inter-subject variability in global excitability. This enables the whole-brain model to replicate individual differences in both g-MS and g-SE, including their covarying relationship, as observed in empirical BOLD signals. For simplicity, both the extents of heterogeneity across the modules ($K^M$) and within each of the seven modules ($K_J^R$, *J*=1, 2, …, 7) were controlled by a single parameter $K$, i. e., $K^M = K_J^R \equiv K$. Our previous study has demonstrated that with in a more simplified case ($T_J^M = 0$), the personalized whole-brain model is already capable of simulating individual differences in resting-state and task-state FC, as well as the extent of updates in FC networks when the brain transitions from resting to task state[45].

*The simulated BOLD signals*: The simulated BOLD signal $x_k(t)$ for the $k$-th ROI were obtained by convolving the point processes $s_k(t)$ generated by the GH automatons with a hemodynamic response function (HRF)[50], i.e.,

$$x_k(t) = \int_0^\infty s_k(t - \tau) h(\tau) d\tau, \tag{11}$$

with

$$h(\tau) = \left(\frac{\tau}{d}\right)^{p-1} \left(\frac{e^{-\frac{\tau}{d}}}{d(p-1)!}\right)^{p-1}, \tag{12}$$

where $d = 0.6$ is the time scaling, and $p = 3.0$ is an integrated phase-delay (the peak delay is given by $pd$, and the dispersion is given by $pd^2$).

**Avalanche statistics and avalanche propagation matrix**

*Avalanche detection*: We conducted avalanche analysis on both computational model outputs and fMRI datasets using spatiotemporally defined neural events. In the model, events were identified by nodal excitation states ("E" states). For the fMRI data, by binarizing the preprocessed BOLD signals at a chosen thresholding (e.g., 2 SD above the mean amplitude, SD: standard deviation) , events were defined by detecting the suprathreshold peak positions intermediate between two above-threshold time points[12, 91]. The above-defined spatiotemporal events were binned with an appropriate temporal resolution (for model it is one iteration step, and for data, it is a TR of 0.72s) and an avalanche was defined as a series of consecutively bins with at least one event, which were led and followed by blank bins without events. The size $S$ and duration $L$ of the avalanches were then defined as the total number of events and total number of time bins during this avalanche, respectively (Fig. 4**a**).

*Avalanche statistics*: The theory of criticality suggests that the distributions of avalanche size $S$ and duration $L$ obey the following power laws[92]:

$$P(S) \sim S^{-\alpha}, \tag{13}$$

$$P(L) \sim L^{-\tau}. \tag{14}$$

Moreover, the mean size $\langle S \rangle$ of avalanches versus a given duration $L$ is governed by another power

law:

$$\langle S \rangle(L) \sim L^{\gamma}. \tag{15}$$

The general scaling theory also predicts the scaling relation among the above three exponents[92]:

$$\frac{\tau - 1}{\alpha - 1} = \gamma. \tag{16}$$

In this study, the truncated power laws were fitted using the NCC toolbox[93], which could automatically fit power laws with left and right cutoffs based on the maximum likelihood estimation. To identify the critical dynamics, we computed the MSD between the cumulative distribution functions of the empirical data and the best-fitting power law. The best-fitting power law was defined by the exponent estimated directly from the empirical distribution. The MSD then quantifies the discrepancy between these two distributions, serving as an indicator for criticality.

**The construction and quantification of the avalanche propagation matrices**

As shown in Fig. 3**a**, for an avalanche, avalanche 2 for example, if there are $M$ nodes excited at iteration step $n$, node $i$ is one of them, and there are $N$ nodes excited at step $n+1$, node $j$ is one of them, then the element $(i, j)$ of the excitation propagation matrix was added $\frac{1}{M} \cdot \frac{1}{N}$. The obtained excitation propagation matrix represents the relative probability for avalanche propagation between any pair of nodes in the network. To make it less skewed so that it can be comparable with the SC and FC matrix, we performed log transformation on the excitation propagation matrix before analysis.

In top panels of Fig. 4**a** and **b**, for better visualization, the avalanche propagation matrices are displayed with their coarse-grained form. Specifically, these matrices were reconstructed by grouping nodes into the above-defined seven modules, where each element is calculated as the average value of the elements within the corresponding seven modules, representing the average propagation strength between pairs of modules. The diagonal elements of the coarse-grained avalanche propagation matrices represent the intra-module propagation strength, reflecting how frequently avalanches propagate within the same module. To analyze disparities in intra-module propagation, we ranked these diagonal values (from lowest to highest) and computed the slope of a linear fit to these ranked values. A steeper slope indicates higher discrepancy in intra-module propagation strength across modules (Fig. 5**a**, middle panels).

The extent of modularity of the avalanche propagation matrices were computed using the same method applied to the SC matrices (Eq. (1)). To visualize the modular structure in the avalanche propagation matrices, we employed the force-directed layouts algorithms[94], which transform abstract networks into visually intuitive maps by intensifying intra-cluster attractions, allowing modules to emerge as tightly connected, coherent groups.

**Measurements for individual cortical FC network properties: segregation, integration, balance between them, flexibility and diversity**

All subsequent analyses were performed on cortical FC networks, from which subcortical nuclei had been excluded.

*Network segregation, integration, and balance between them:* We applied the nested-spectral partition (NSP) method[11] to quantify the segregation and integration ability of FC networks using eigenmodes. Eigen-decomposition of the FC matrix yielded eigenvectors U and eigenvalues $\Lambda$, with hierarchy levels defined by eigenvalue magnitude. Hierarchical modules were identified by partitioning eigenvector components into positive/negative signs at each level, iteratively subdividing modules from the first level onward. Module counts $M_i$ $(i = 1, \dots, N)$ and sizes $m_j$ $(j = 1, \dots, M_i)$ were tracked across levels. The segregation and integration components at each level were computed as[11]:

$$H_i = \frac{\Lambda_i^2 M_i (1 - p_i)}{N},$$ (17)

where $p_i = \sum_j |m_j - N/M_i|/N$. Here, $N$ is the number of brain regions, and $p_i$ corrects for modular size deviations. The global integration component was then derived from the first level:

$$\widetilde{H}_{In} = \frac{H_1}{N},$$ (18)

while segregation $H_{Se}$ summed contributions from levels 2 to N:

$$\widetilde{H}_{Se} = \sum_{i=2}^{N} \frac{H_i}{N}.$$ (19)

Then, the segregation-integration balance was defined as $\widetilde{H}_B = \widetilde{H}_{In} - \widetilde{H}_{Se}$.

Because the shorter length of an fMRI series biased the network to more segregation[11, 95], we constructed the stable FC network by concatenating all fMRI time series of all participants in each group, and estimated the group-stable segregation $H_{Se}^S$ and integration $H_{In}^S$ components that were calibrated to the corresponding values of the stable FC. For individual participants, their integration $H_{In}$ and segregation $H_{Se}$ were calibrated to $H_{In} = \widetilde{H}_{In} \times H_{In}^S / \langle H_{In} \rangle$ and $H_{Se} = \widetilde{H}_{Se} \times H_{Se}^S / \langle H_{Se} \rangle$. Here, $\widetilde{H}_{In}$ and $\widetilde{H}_{Se}$ are the individual participant's integration and segregation measures, and $\langle \dots \rangle$ represents the average of $\widetilde{H}_{In}$ and $\widetilde{H}_{Se}$ across all participants. The generated the fMRI length-independent network measures for all participants in each group and has been found to be advantageous in linking the brain to cognitive abilities[95].

*FC flexibility*: FC flexibility quantifies the frequency of transitions between segregated and integrated states in an individual's FC dynamics[11]. This measure was derived by first applying a sliding window (length: 83 TRs, step: 1 TR) to the BOLD signals to generate windowed FC matrices. From these, we derived a time-dependent balance measure, $H_b(t)$, between temporal segregation and integration for each participant. Specifically, to maintain the individual rank, for each participant, the temporal segregation and integration components were calibrated to the respective components computed from the full-length FC time series[9]. FC flexibility was then defined as the proportion of time points where a state transition occurred, indicated by a sign change in $H_b(t)$, relative to the total scanning time $t_{all}$:

$$F_{FC} = \frac{n_{H_b(t)H_b(t+1) \leq 0}}{t_{all}}, \qquad (20)$$

where $n$ is the total number of switching times that satisfied $H_b(t)H_b(t+1) \leq 0$, and $t_{all} = 4717$ TR.

*FC diversity:* The FC diversity characterizes the broadness of the distribution of the FC values in the FC network and is defined as similarity between the empirical FC distribution and a theoretical uniform distribution[9, 96]:

$$D_{FC} = 1 - \frac{1}{N_M} \sum_{i=1}^{M} \left| p_i - \frac{1}{M} \right|, \qquad (21)$$

where $N_M = 2\frac{M-1}{M}$ is a normalization factor, and $p_i$ is the probability of the elements in FC matrix elements falling into the $i$-th bin, with the interval $[0,1]$ discretized into $M = 40$ bins. $D_{FC} \approx 1$ signifies uniform distribution of FC elements (high diversity), whereas $D_{FC} \approx 0$ reflects confinement to few bins (low diversity).

**SEM-based estimation of the latent factors for general and three domain-specific cognitive abilities**

As in the previous study[12]，we included nine cognitive tasks listed in the HCP Data Dictionary: picture sequence memory (PSM), dimensional change card sort (DCCS), flanker inhibitory control and attention task (FT), Penn progressive matrices (PPM), oral reading recognition (ORR), picture vocabulary (PV), pattern completion processing speed (PCPS), variable short Penn line orientation test (VSPLOT), and Penn word memory test (PWMT). We employed a multiple-group bifactor minus-1 SEM model to extract latent factors representing common phenotypes, which account for the variability observed across the aforementioned tasks. Multiple-group models assume that the same factor structure and structural relationships apply across different groups while also allowing for the empirical testing of potential nonlinear associations between variables across these groups. The subjects were divided into subcritical, GP, and supercritical groups based on the criteria outlined above, with the GP group further subdivided into a few approximately equal size groups. The multiple-group model, comprising general cognitive ability and three group-specific factors, demonstrated a good fit to the data (CFI = 0.977; RMSEA = 0.044; SRMR = 0.035). The three factors are represented as: (1) Crystallized intelligence (ORR + PV); (2) Processing speed (DCCS + FT + PCPS); and (3) Memory (PWMT + PSM). SEM analysis was performed using the *lavaan* package in R[97].

**Acknowledgements**

Wu and Yu were supported by the National Natural Science Foundation of China (Grant No. 12247101), the Fundamental Research Funds for the Central Universities (Grant No. lzujbky-2021-62 and lzujbky-2024-jdzx06), the Natural Science Foundation of Gansu Province (No. 22JR5RA389), and the '111 Center' under Grant No. B20063. Zhou was partially supported by the Hong Kong Research Grant Council (RGC) Senior Research Fellow Scheme (SRFS2324-2S05) and RGC General Competitive Fund (GRF12202124), Guangdong and Hong Kong Universities "1+1+1" Joint Research Collaboration Scheme (2025A0505000011) and Hong Kong Baptist University Seed Funding for Collaborative Research Grants (RC-SFCRG/23-24/SCI/06).

**Data availability**

The human brain imaging data used in this study are publicly available through the Human Connectome Project (HCP) database (https://www.humanconnectome.org). The scripts used in this study are available from the corresponding author upon reasonable request.

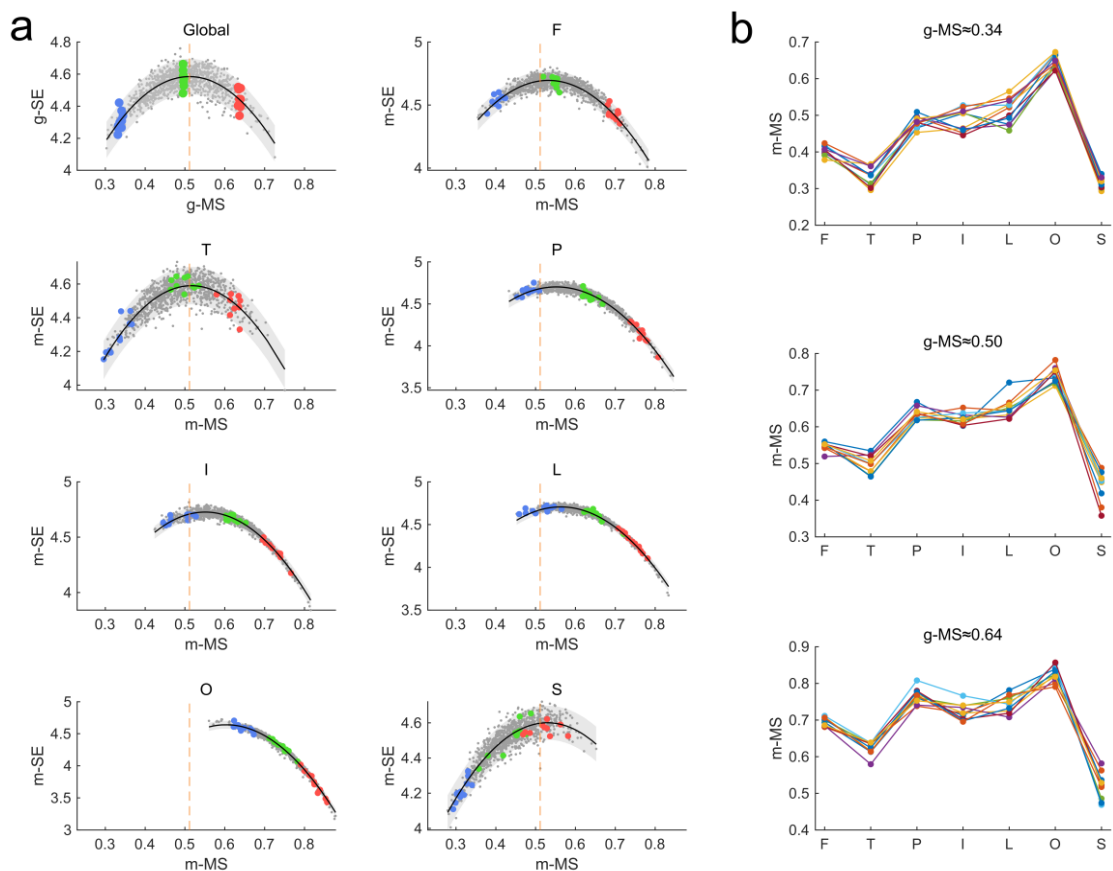

**Fig. S1 The individual variability in local dynamics with identical global network dynamics. a**: Three participant groups (10 subjects in each group) with nearly identical g-MS values (colored dots in global plot: blue: g-MS≈0.34；green：g_MS≈0.50；red: g-MS≈0.64) exhibit divergent local dynamics (indicated by spread m-MS in seven modular m-MS *vs.* m-SE plots). **b**: Detailed view of the modular m-MS values for the three groups (F=frontal, T=temporal, P=parietal, I=insular, L=limbic, O=occipital, S=subcortical). For better visualization, only data for 10 subjects are presented.

**Table S1. List of the 7 modules with their anatomical regions (BN246 atlas).**

| ID for 7 modules | ID for ROIs | Gyrus | Lobe |
|---|---|---|---|
| 1 | 1~14 | SFG, Superior Frontal Gyrus | Frontal Lobe |
| | 15~28 | MFG, Middle Frontal Gyrus | |
| | 29~40 | IFG, Inferior Frontal Gyrus | |
| | 41~52 | OrG, Orbital Gyrus | |
| | 53~64 | PrG, Precentral Gyrus | |
| | 65~68 | PCL, Paracentral Lobule | |
| 2 | 69~80 | STG, Superior Temporal Gyrus | Temporal Lobe |
| | 81~88 | MTG, Middle Temporal Gyrus | |
| | 89~102 | ITG, Inferior Temporal Gyrus | |
| | 103~108 | FuG, Fusiform Gyrus | |
| | 109~120 | PhG, Parahippocampal Gyrus | |
| | 121~124 | pSTS, posterior Superior Temporal Sulcus | |
| 3 | 125~134 | SPL, Superior Parietal Lobule | Parietal Lobe |
| | 135~146 | IPL, Inferior Parietal Lobule | |
| | 147~154 | Pcun, Precuneus | |
| | 155~162 | PoG, Postcentral Gyrus | |
| 4 | 163~174 | INS, Insular Gyrus | Insular Lobe |
| 5 | 175~188 | CG, Cingulate Gyrus | Limbic Lobe |
| 6 | 189~198 | MVOcC, MedioVentral Occipital Cortex | Occipital Lobe |
| | 199~210 | LOcC, lateral Occipital Cortex | |
| 7 | 211~214 | Amyg, Amygdala | Subcortical Nuclei |
| | 215~218 | Hipp, Hippocampus | |
| | 219~230 | BG, Basal Ganglia | |
| | 231~246 | Tha, Thalamus | |

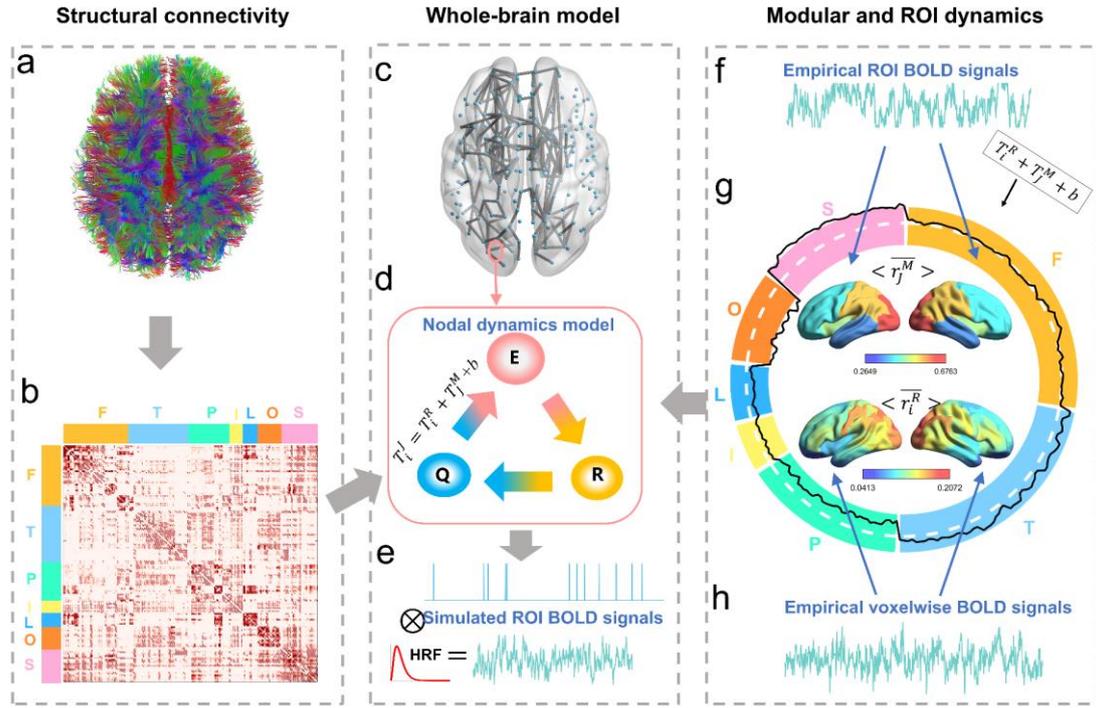

**Fig. S2 The construction of the whole-brain model and the calibration for the regional heterogeneity.** The whole-brain model was constructed as follows (**a-e**): The tractography of DTI (**a**) between the 246 ROIs according to the BN246 parcellation was processed to derive the group-averaged SC matrix (**b**), based on which the whole-brain network model was constructed (**c**). In the model, the nodal dynamics was described by the GH automaton (**d**), which means each node in the network could take one of the three states: quiescent (Q), excitable (E), and refractory (R). Every node updates its state according to the transition rule (Methods). Specifically, the $i$-th node in the $J$-th module would transit from "Q" to "E" if it received inputs exceeding its excitation threshold $T_i^J$. The whole-brain model generates simulated BOLD signals for each ROI by convolving the converted binary point process (Q/R states = 0, E states = 1; **e**, top) with a standard hemodynamic response function (**e**, bottom). The heterogeneous nodal excitation threshold $T_i^J$ was calibrated with both modular and regional synchronization dynamics in the following way (**f-g**): $<\overline{r_J^M}>$, the m-MS of the $J$-th module (top panels inside the circles of (**g**)) was calculated from the empirical BOLD signals of ROIs in this module (**f**). Additionally, $<\overline{r_i^R}>$, the MS of the $i$-th ROI within the $J$-th module (bottom panels inside the circles of **g**) was calculated from the voxel-wise BOLD signals of voxels in this ROI (**h**). With the proposed calibrating approach (Methods), $T_i^R$, the threshold for the $i$-th ROI, and $T_J^M$, the threshold for the $J$-th module, were determined by $<\overline{r_i^R}>$ and $<\overline{r_J^M}>$, respectively (Eqs. (9) and (10)). Finally, $T_i^J$, the threshold for each node in **d**, were determined as $T_i^J = T_i^R + T_J^M + b$ for a given extent of heterogeneity $K$, where $b$ is the baseline threshold (Eq. (8)). An example ($b$=0 and $K$=6) of the nodal threshold distribution in each module was demonstrated with black line in the circle of **g**, where the white dashed line indicates the mean vale of these thresholds. In **b** and **g**, F=frontal, T=temporal, P=parietal, I=insular, L=limbic, O=occipital, and S=subcortical.

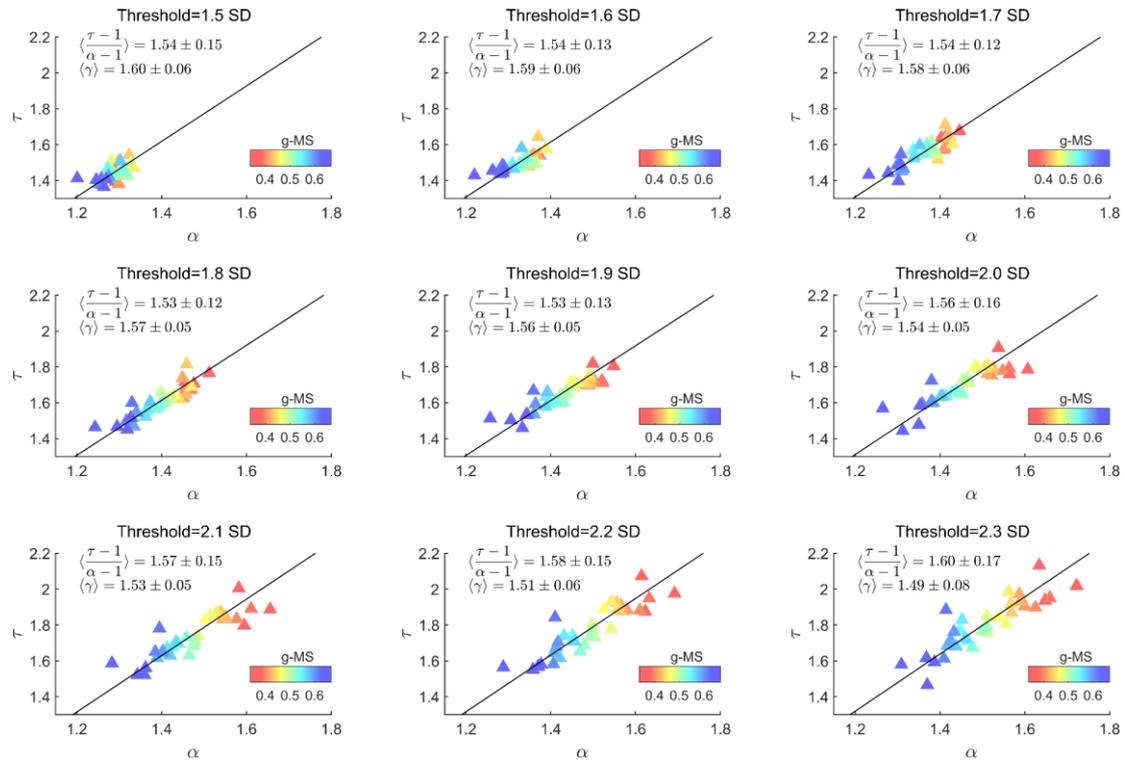

**Fig. S3 The robustness of the scaling relation calculated from the fMRI data with respect to different binarizing thresholds.** Individuals were ranked by g-MS and grouped into groups of 30 subjects with comparable g-MS values for calculating avalanche exponents and group-mean g-MS. Each triangle denotes one group, with color encoding its group-mean g-MS value as indicated by the color bar.